\documentstyle[12pt]{article}
\newcommand{\be}{\begin{equation}}
\newcommand{\bea}{\begin{eqnarray}}
\newcommand{\eea}{\end{eqnarray}}
\newcommand{\ba}{\begin{array}}
\newcommand{\ea}{\end{array}}
\newcommand{\ee}{\end{equation}}

\expandafter\ifx\csname mathbbm\endcsname\relax

\else

\fi
\begin{document}

\begin{titlepage}
\begin{flushright}
IP/BBSR/2002-19\\
IPM/P-2002/030\\
hep-th/0207257
\end{flushright}

\vspace{5mm}
\begin{center}
{\Large {\bf PP-waves from Nonlocal Theories}\\}
\vspace{16mm}

{Mohsen Alishahiha$^a$\footnote{alishah@theory.ipm.ac.ir} 
and Alok Kumar$^b$\footnote{kumar@iopb.res.in}}\\
\vspace{2mm}
{\em $^a$ Institute for Studies in Theoretical Physics and Mathematics (IPM)\\
P.O. Box 19395-5531, Tehran, Iran\\ }
\vspace{2mm}
{\em $^b$ Institute of Physics\\ 
Bhubaneswar 751 005, India\\}
\vspace{2mm}
\end{center}

\begin{abstract}
We study the Penrose limit of ODp theory. There are two different
PP-wave limits of the theory. One of them is a ten dimensional PP-wave 
and the other a four dimensional one. We observe the later one leads 
to an exactly solvable background for type II string theories where
we have both NS and RR fields.
The Penrose limit of different branes of string (M-theory) in a nonzero 
B/E field (C field) is also studied. These backgrounds are conjectured to 
provide dual description of NCSYM, NCOS and OM theory. We see that under 
S-duality the subsector of NCSYM$_4$ and NCOS$_4$ which are dual to the 
corresponding string theory on PP-wave coming from  NCYM$_4$ and NCOS$_4$ 
map to each other for given null geodesic. 
 
\end{abstract}
\end{titlepage}

\newpage

\section{Introduction}

Recently, PP-wave backgrounds in string theory have attracted 
much interest, since they possess several interesting features, 
such as exact solvability in  
Green-Schwarz formalism etc.. In fact an
interesting observation in \cite{metsaev}
is that type IIB string theory on the 10-dimensional PP-wave 
background with constant five-form RR field strength is exactly solvable. 
Moreover it has also been shown \cite{hull} that this background is 
a maximally supersymmetric solution of type IIB supergravity.

Another interesting fact is that such a $D=10$ maximally
supersymmetric PP-wave background can be obtained as a Penrose
limit of another one with maximal supersymmetry, namely
$AdS_5\times S^5$ \cite{{blau},{BMN}}. On the other hand type IIB
string theory on $AdS_5\times S^5$ is believed to be dual to
the ${\cal N}=4$ SYM theory in four dimensions. Indeed it has
been shown \cite{BMN} that taking Penrose limit of
$AdS_5\times S^5$ has a corresponding limit for the gauge theory
as well. By making use of this fact the authors of \cite{BMN}
have been able to identify the excited string states with a
class of gauge invariant operators in the large $N$ limit of
${\cal N}=4$ SYM $SU(N)$ theory in four dimensions. Such an
identification is very difficult for $AdS_5\times S^5$
itself mainly because the string theory in this maximally
supersymmetric background has not proved to be exactly solvable
yet.

Considering the fact that the string theory on a PP-wave background,
obtained by taking the Penrose limit of $AdS_5\times S^5$, is
exactly solvable one might wonder if this is the case for other
gravity backgrounds of string theory. Therefore it would be 
important to study the Penrose limit of different known supergravity
solution in string theory. Actually soon after \cite{BMN}, several 
papers appeared discussing Penrose limit of different gravity
solutions in string theory \cite{all2} (see also \cite{all1}). 

Recently the Penrose limit of supergravity solutions which 
are supposed to be
dual to some non-local theories have been studied in \cite{HRV}, where
the authors considered little string theory, (0,2) theory and four 
dimensional noncommutative gauge theory \footnote{For definition of these 
theories see \cite{Berkooz}-\cite{MalRU}.}. These theories are conjectured to 
be dual to the theories on the worldvolume of NS5-brane in type II string 
theories, M5-branes and D3-brane with B field in their decoupling limit,
respectively. 

The aim of this paper is to further study the PP-waves from non-local
theories. The theories we shall consider is ODp-theory, NCOS and OM
theory \cite{{OM},{GMMS},{HARM1}}. The corresponding supergravity 
solutions have also been studied in \cite{HARM1}-\cite{all5}.

Our work is also motivated by a recent analysis of ``time dependent''
PP-wave background \cite{RG} (see also \cite{RG1,CAS}), 
where particle motion has been shown to be exactly
solvable, with fixed points of the Hamiltonian being related to 
the time-independent harmonic motion of massive particles. In addition, 
non-constant `masses' appear in the Green-Schwarz worldsheet theory
and turn out to be of importance in undestanding the
RG flow in the corresponding gauge theory. More precisely, 
different fixed points of the renormalization group flow
in guage theory side, give rise to different expressions for 
masses and thereby determine the validity of the supergravity 
description. In this paper we will also present several examples of
non-constant `time dependent', masses and discuss their implications
for our solutions. 

The organization of the paper is as following. In section 2 we
will study the Penrose limit of ODp theory. We shall see that
there are two different limits, one of which leads to a
background in which the string theory can be exactly solved.
In section 2 the Penrose limit of NCOS$_4$ is considered where
we observe that the resulting PP-wave is S-dual to one coming
from NCSYM$_4$. In section 4 we shall consider the Penrose limit
of OM-theory. The last section is devoted to the conclusions and
comments.

\section{Penrose limit of type II NS5-branes in the presence of RR field}

There are decoupled theories (ODp) on the worldvolume of 
type II NS5-branes in the presence of nonzero $p$-form RR filed
\cite{{OM},{HARM1}}.\footnote{For $p=1,2$ see also \cite{ALI1}.}
The excitations of these theories include light open Dp-branes.
The gravity description of these theories have been studied in
\cite{AOR,mitra-roy}.

In this section we will consider the Penrose limit of these
theories. We shall consider two different null geodesics
for these backgrounds leading to two PP-wave solutions
in type II string theories.

The gravity description of ODp theory is given by 
\bea
l_s^{-2}ds^2&=&(1+a^2r^2)^{1/2}\left[-dt^2+\sum_{i=1}^{p}
dx_{i}^2+\frac{\sum_{j=p+1}^{5}d x_{j}^2}{1+a^2r^2}
+{ N\over r^2}(dr^2+r^2d\Omega^2_3)\right]\ ,\cr
&&\cr
A_{0\cdots p}&=&{l_s^{(p+1)} \over {\tilde g}} a^2r^2\ \ ,
\;\;\;\;\;\;\;\;\;\;
\;\;\;\;\;\;
A_{(p+1)\cdots 5}={l_s^{(5-p)}\over {\tilde g}}\;\frac{a^2r^2}{1+a^2r^2}\ ,
\cr
&&\cr
e^{2\phi}&=&{\tilde g}^2 \frac{(1+a^2r^2)^{(p-1)/2}}{a^2r^2},
\;\;\;\;\;\;\;l_s^2dB=2N\;\epsilon_3,\;\;\;\;a^2={l_{\rm eff}^2\over N}\;,
\label{rrrr}
\eea
where $l_{\rm eff}$ and ${\tilde g}$ are effective string tension and 
effective string coupling of the theory which are the parameters of 
the theory after taking the decoupling limit \cite{OM}. 
Here $\epsilon_3$ is the volume of $S^3$ part of the metric.

To study the Penrose limit of the above gravity solution we will first rescale
$t\rightarrow \sqrt{N} t$ and consider a null geodesic 
in the $(t,r,\psi)$ plane at the fixed point with
respect to other coordinates\footnote{Here we parameterize the 3-sphere
in (\ref{rrrr}) as $d\Omega_3^2=d\theta^2+\cos^2\theta\;d\psi^2
+\sin^2\theta\;d\beta^2$.}. This geodesic is generated by tangent vector 
${\dot t}\partial_t+{\dot r}\partial_r+{\dot \psi}\partial_{\psi}$, where
dot denotes derivative  with respect to the affine parameter.
Since $\partial_t$ and $\partial_{\psi}$ are killing vectors, they define
constant of motion along geodesic. In fact, defining $h=1+a^2r^2$,
\be
E=h^{1/2}{\dot t},\;\;\;\;\;\;J=h^{1/2}{\dot \psi}
\ee
are conserved quantities corresponding to the energy and angular momentum.
For a null geodesic we get
\be
{\dot r}^2={r^2\over h}(1-l^2),
\label{NULL}
\ee
where $l={J\over E}$ and we have also rescaled the affine parameter by $E$. One
could solve this differential equation to find $u$ as a function of $r$.  
This could be inverted to find $r(u)$ and then we can plug this into the other 
expressions in the metric to write the metric in terms of $u$.

Consider a coordinate transformation from $(t,r,\psi)\rightarrow
(u,v,x)$ which is more suitable coordinate to write the metric. The 
coordinate transformation is given by
\bea
dr&=& {r\over \sqrt{h}}(1-l^2)^{1/2}du\;,\cr &&\cr
dt&=&{1\over \sqrt{h}}du-dv+ldx\;,\cr &&\cr
d\psi&=&{l\over \sqrt{h}}du+dx\;,
\eea
Substituting this change of coordinate in the metric
 and rescaling the coordinate as following
\be
u\rightarrow u,\;\;\;v\rightarrow{v\over N},\;\;\;\theta\rightarrow
{z\over \sqrt{N}},\;\;\;\;x\rightarrow{x\over \sqrt{N}},
\ee 
with all other coordinates remaining fixed, and then taking the Penrose
limit ($N\rightarrow \infty$) we find the PP-wave limit metric of
ODp-theory as following
\bea
ds^2&=&2dudv-{l^2\over \sqrt{h}}z^2du^2+h^{1/2}\bigg{(}
(1-l^2)dx^2+d{\vec z}^2\cr
&+&dy^2+y^2d\Omega_{p-1}^2 \bigg{)}+h^{-1/2}(dw^2+w^2d\Omega_{4-p}^2)
\eea
in this metric $h$ is a function of $u$. This is the PP-wave in Rosen 
coordinate.

To find the PP-wave in  Brinkman we use the following transformation
\be
u\rightarrow u,\;\;\;\; x\rightarrow {x\over h^{1/4} \sqrt{1-l^2}},\;\;\;
z\rightarrow {z\over  h^{1/4}},\;\;\;
\;\;\;y\rightarrow {y\over h^{1/4}},\;\;\;\;w\rightarrow h^{1/4}w
\ee
and
\be
v\rightarrow v- {1\over 2}\frac{\partial_u h^{1/4}}{h^{1/4}} (x^2+y^2+z^2
-w^2)
\ee
Using this transformation the metric reads
\be
ds^2=2dudv+\left(F_1\;(x^2+ y^2+ z^2)+F_2w^2-{l^2\over h}z^2
\right)du^2+dx^2+
d{\vec y}^2+d{\vec z}^2+d{\vec w}^2,
\ee
where
\be
F_1=\frac{\partial_u^2  h^{1/4}}{h^{1/4}}\;,\;\;\;\;\;\;\;\;
\;\;
F_2=2\left(\frac{\partial_u h^{1/4}}
{h^{1/4}}\right)^2-\frac{\partial_u^2  h^{1/4}}{h^{1/4}}\;.
\ee

We note, however, that this is not the whole story. We need to know which
fields survive in this limit. In fact there is a free parameter in the theory
whose scaling behavior is to be taken into account. Namely, we need to know 
the behavior of effective string length in the large $N$ limit. Indeed
keeping $l_{\rm eff}$ fixed in large $N$ limit means that $a\rightarrow
0$ and therefore $h\rightarrow 1$. In this case, for the fixed scale where we 
have $l=1$, we just get a PP-wave which is the same as if we had considered 
the Penrose limit of NS5-branes, so we get
\bea
ds^2&=&-2dudv-\mu^2z^2du^2+d{\vec z}^2+d{\vec Y}_6^2\;,\cr 
H_{uz_1z_2}&=&2\mu\;,
\label{NSPP}
\eea
with constant dilaton. Here we have properly rescaled $du$ and $dv$ 
by $\mu$. This is exactly the 
NS PP-wave which was derived in \cite{NW}. This background represents
an exact string background to all orders in the worldsheet theory 
\cite{{Amati},{HOR}}. This background has also been studied in
\cite{KP} (see also \cite{HRV}). It is also easy to see that the
solution (\ref{NSPP}) does solve supergravity equations of motion.
To see this we note that the $\mu$ dependence
appears nontrivially only in the Ricci tensor $R_{++}$, where one
obtains a constant contribution of $2\mu^2$, which is canceled by
$2\mu^2$ contribution coming from B field.


We note, however, that since in the solution (\ref{rrrr}) $a$ always appears 
in the combination of $ar$ it is natural to consider a case in which the 
effective deformation parameter \cite{AOS}, $ar$,  remains fixed in the 
Penrose limit. In this case, setting $ar=e^{\rho}$, the PP-wave
obtained  from the null geodesic (\ref{NULL}) becomes 
\bea
ds^2&=&2dudv+\left(F_1\;(x^2+ y^2+ z^2)+F_2w^2-{l^2\over h}z^2
\right)du^2\cr &&\cr 
&+&dx^2+d{\vec y}^2+d{\vec z}^2+d{\vec w}^2,\cr
&&\cr
e^{2\phi}&=&{\tilde g}^2 \frac{(1+e^{2\rho})^{(p-1)/2}}{e^{2\rho}}\;,
\;\;\;\;\;\;\;\;\;\;
\;\;\;\;\;\;
A_{(p+1)\cdots 5}={l_s^{(5-p)}\over {\tilde g}}\;\frac{e^{2\rho}}
{1+e^{2\rho}}\;,\cr &&\cr
dA^{(p+1)}&=&{2l(1-l^2)^{1/2}\over {\tilde g}}\;{e^{2\rho}\over 
\sqrt{1+e^{2\rho}}}\;dx\wedge du \wedge dx_1\wedge\cdots \wedge dx_p\;,
\cr &&\cr
dB&=&{2l\over \sqrt{1+e^{2\rho}}}\;du\wedge dz_1\wedge dz_2\;,
\eea
where $F_1$ and $F_2$ are now given by
\be
F_1={(1-l^2)\over 4}\;{e^{2\rho}(4-e^{2\rho})\over (1+e^{2\rho})^3}\;,
\;\;\;\;\;\;\;\;\;\;
F_2={(1-l^2)\over 4}\;{e^{2\rho}(3e^{2\rho}-4)\over (1+e^{2\rho})^3}\;.
\ee
Here $\rho$ is a function of $u$ given by
\be
\sqrt{1+e^{2\rho}}-\tanh^{-1}\sqrt{1+e^{2\rho}}=(1-l^2)u\;.
\ee
It can be shown that this PP-wave solves the corresponding type II 
supergravity equations of motion.

Since the PP-wave we found is time dependent ($u$-dependent)
one can study the RG flow of the theory. For example,
from point of view of the 2-dimensional quantum theory on string 
worldsheet the masses of fields are given by $-F_1$ and $-F_2$. 
But from the expression of $F_1$ and $F_2$, we observe that in UV or 
IR some directions get imaginary masses. This means the gravity 
description we started with is not a good one and we need to change 
our description we can be done either using S-duality 
(in type IIB) or lifting the theory to M-theory (in type IIA).

We now present
another plane wave solution which is obtained by taking 
Penrose limit corresponding to a longitudinal null geodesic which also 
leads to an exactly solvable string theory independent of what $a$ is. 
For this, we perform a rescaling of $t$ and $x_i$, $(i=1,..,5)$ by a factor 
$\sqrt{N}$ and make a coordinate transformation: $ a r = e^{\rho}$. 
Then the metric as well as other fields in (\ref{rrrr}) read:
\bea
ds^2&=& l_s^2 N h^{1/2}\bigg{[}-dt^2+\sum_{i=1}^{p}
dx_{i}^2+ h^{-1} {\sum_{j=p+1}^{5}d x_{j}^2} +d\rho^2 \cr &&\cr
&&\;\;\;\;\;\;\;\;\;\;\;\;\;\;
+ d\theta^2 + \sin^2\theta\; d\beta^2 + \cos^2\theta\; d\psi^2
\bigg{]}\ ,\cr
&&\cr
A_{0\cdots p}&=&{l_s^{(p+1)}  \over {\tilde g}}N^{(p+1)/2} e^{2\rho}\ \ ,
\;\;\;\;\;\;\;\;\;\;
A_{(p+1)\cdots 5}=
{l_s^{(5-p)}\over {\tilde g}}\;\frac{N^{(5-p)/2}e^{2\rho}}{1+ e^{2\rho}}\ ,
\cr &&\cr
e^{2\phi}&=&{\tilde g}^2 \frac{(1+ e^{2\rho})^{(p-1)/2}}{e^{2\rho}},\;\;\;\;\;
\;\;\;\;\;\;
l_s^{-2}dB=2N\epsilon_3\;.
\label{rrrr2}
\eea 
Now consider a null geodesic around $\rho=\rho_0$=constant. In this case
we perform the following coordinate transformation:
\be\ba {ll}
\rho =\rho_0 + ( 1 + e^{2\rho_0})^{-1/4}{r\over \sqrt{N}}, &
x'_j=( 1 + e^{2\rho_0})^{1/4}{w_j\over \sqrt{N}},\;\;(j=p+1,..,5),\cr
\theta =( 1 + e^{2\rho_0})^{-1/4}{z \over \sqrt{N}}, &
x_i = ( 1 + e^{2\rho_0})^{-1/4}{y_i\over \sqrt{N}}\;\;(i=1,..,p),\cr
t =( 1 + e^{2\rho_0})^{-1/4}(x^+ + {x^-\over N}),&
\psi = ( 1 + e^{2\rho_0})^{-1/4}(x^+ - {x^-\over N})\;,
\ea\ee
and $\beta$ is fixed. One then obtains in the Penrose limit 
($N\rightarrow \infty$) a PP-wave background as following
\be
ds^2 = -4 dx^+ dx^- - \mu^2 z^2 {dx^+}^2 + d{\vec z}^2+d{\vec Y}^2_6\;,
\label{RNSPP}
\ee
where $d{\vec Y}^2_6=dr^2+d{\vec w}^2+d{\vec y}^2$.
Here we have also rescaled the longitudinal coordinates by
$x^{\pm}\rightarrow x^{\pm} (1 + e^{2\rho_0})^{\pm{1/4}}$. 
In this limit, one gets also a nonvanishing $p+1$ form field and nonzero
B field:
\bea
dA^{(p+1)} &=& {2\mu\over \tilde{g}}\; e^{2\rho_0} (1+ e^{2\rho_0})^{-(p+1)/4}
              \; dx^+ \wedge dr \wedge dy_1 \wedge...\wedge dy_p\;,\cr &&\cr
A_{(P+1)\cdots 5}&=&{1\over {\tilde g}}\;\frac{e^{2\rho_0}}{1+e^{2\rho_0}}
\;dw_{(P+1)} \wedge\cdots \wedge dw_5 \;,\cr &&\cr
dB&=&2\mu(1+e^{2\rho_0})^{-1/2} \;dx^+\wedge dz_1\wedge dz_2\;,
\label{RNSPPF}
\eea
and the constant dilaton after the Penrose limit is given by:
\be
e^{2\phi} = \tilde{g}^2 {(1 + e^{2 \rho_0})^{(p-1)/2}\over
e^{2\rho_0}}\;.
\ee
Here we have properly rescaled $dx^+$ and $dx^-$ 
by $\mu$. One can again check that this solution solves the type II
supergravities
equations of motion. To see this we note that the $\mu$ dependence
appears nontrivially only in the Ricci tensor $R_{++}$, where one
obtains a constant contribution of $2\mu^2$, which is canceled by
$2\mu^2(1+e^{2\rho_0})^{-1}$ contribution coming from B field
 and $2\mu^2 e^{2\rho_0}(1+e^{2\rho_0})^{-1}$
contribution from RR (p+1)-form.

We have therefore obtained a four dimensional PP-wave background
from ODp theories. The form of the background is independent of the
value of $p$ one started with. The metric obtained for this background is
very similar to that of NS PP-wave (\ref{NSPP}), though here we also have
nonzero RR field. One can show that these backgrounds also correspond to 
exactly solvable string theories.  

To see this one can write down the GS action for the superstring in this
background. In the light-cone gauge the bosonic action leads to two 
massive fields ($z_1,z_2$) and six free scalar fields ${\vec Y}_6$. 
The fermion mass comes from the RR field and 
is dependent on the matrix $\Pi \equiv \gamma^1...\gamma^p$.

To be precise the bosonic 2-dimensional worldsheet action in the 
presence of non-zero B-field is given by
\be
S_{\rm bos}=-{1\over 4\pi\alpha'}\int
d\sigma^2\left[\eta^{ab}G_{\mu\nu}
\partial_a x^{\mu}\partial_b x^{\nu}+\epsilon^{ab}B_{\mu\nu}
\partial_a x^{\mu}\partial_b x^{\nu}\right],
\ee
where $\eta={\rm diag}(-1,1)$ is the worldsheet metric and we use a
notation in which $\epsilon^{\tau\sigma}=1$.

Plugging the R-NS PP-wave (\ref{RNSPP}) and (\ref{RNSPPF}) in this action, one
finds
\bea
S_{\rm bos}&=&-{1\over 4\pi \alpha'}\int
d\sigma^2\bigg{[}\eta^{ab}\left( 2\partial_a x^+\partial_b
x^-+\partial_a z_{i}\partial_b z_{i}-
\mu^2 z_i^2\partial_a x^{+}\partial_b x^{+}+\partial_a {\vec Y}
\partial_b {\vec Y}\right)\cr
&&\cr &+&2\mu (1+e^{2\rho_0})^{-1/2}z_2\epsilon^{ab}
\left(\partial_a x^{+}\partial_b
z_{1}-\partial_a z_{1}\partial_b x^{+}\right)\bigg{]}\;,
\label{PPAA}
\eea
where we have rescaled $x^+$ by $\mu$. The action can be simplified using the 
light-cone gauge. In this gauge, setting,
$x^+=p^+\tau$, the action (\ref{PPAA}) reads
\be
S_{\rm bos}=-{1\over 4\pi \alpha'} \int
d\sigma^2\left(\eta^{ab}\partial_a z_i\partial_b z_i+
m^2 z_i^2+ \partial_a {\vec Y}
\partial_b {\vec Y}+{4 m\over \sqrt{1+e^{2\rho_0}}}
z_2\partial_{\sigma} z_1\right),
\label{abos}
\ee
where $m=\mu p^+$. The equations of motion are given by
\bea
&&\eta^{ab}\partial_a\partial_b {\vec Y}=0\;,\cr &&\cr 
&& \eta^{ab}\partial_a\partial_b z_2-m^2
z_2-{2m\over \sqrt{1+e^{2\rho_0}}}\; \partial_{\sigma}z_1=0,\cr &&\cr
&&\eta^{ab}\partial_a\partial_b z_1-m^2 z_1+ 
{2m\over \sqrt{1+e^{2\rho_0}}}\;\partial_{\sigma}z_2=0
\eea
subject to the following boundary conditions
\be
\partial_{\sigma}{\vec Y} \delta {\vec Y}|_{\rm boundary}=0,\;\;\;
\partial_{\sigma}z_2\delta z_2|_{\rm boundary}=0,\;\;\;
\partial_{\sigma}z_1\delta z_1-{2m \over \sqrt{1+e^{2\rho_0}}}\;z_2
\delta z_1|_{\rm boundary}=0\,.
\label{bond}
\ee
Now we can proceed to solve the equation of motion subject to the
above boundary condition. To do this one considers an ansatz as 
$z_i=\alpha_ie^{i(\omega_n \tau+n\sigma )}$ and
${\vec Y}={\vec \beta}e^{i(\omega_n \tau+n\sigma )}$. Then the bosonic
frequencies are given by
\be\ba {ll}
\omega_n=\pm\sqrt{m^2+n^2\pm 2mn (1+e^{2\rho_0})^{-1/2}},\;\;\;\;
&{\rm for}\;\;z_1,z_2\cr 
\omega_n =\pm n,\;\;\;\; & {\rm for}\;\; {\vec Y}\;.
\ea\ee
The general form of the solution is once again similar to the 
that discussed in the literature \cite{metsaev} 
(see also \cite{RT}-\cite{AGGP}).  

Following \cite{metsaev} one can write the quadratic fermionic 
action as following
\be
{i\over\pi}\int d^2\sigma\left(\eta^{ab}\delta_{pq}
-\epsilon^{ab}(\sigma_3)_{pq}\right)\partial_a
x^{\mu}{\bar \theta}^p\Gamma_{\mu}{\cal D}_b\theta^q, \label{FAC}
\label{ACF11}
\ee
where $\theta^p, \; p=1,2$ are two 10 dimensional spinors with
same/different chiralities in type IIB/A, and  $\sigma_3={\rm diag}(1,-1)$. The
generalized covariant derivative, ${\cal D}$ is given by
\cite{{ROM},{Hassan}}
\bea
{\cal D}_a&=&\partial_a+{1\over 4}\partial_a x^{\rho}
\bigg{[}\bigg{(}\omega_{\mu\nu\rho}- {1\over 2} H_{\mu\nu\rho}\sigma_3
\bigg{)}
\Gamma^{\mu\nu}\cr &&\cr &&
\;\;\;\;\;\;\;\;\;\;\;\;\;\;\;\;\;\;\;+
\bigg{(}{1\over 2!} F_{\mu\nu}\Gamma^{\mu\nu}\sigma_0+{1\over 2\times
 4!} F_{\mu\nu\lambda\delta}
\Gamma^{\mu\nu\lambda\delta}\sigma_1\bigg{)}\Gamma_{\rho}\bigg{]},
\label{ACF22}
\eea
for type IIA string theory, and
\bea
{\cal D}_a&=&\partial_a+{1\over 4}\partial_a x^{\rho}
\bigg{[}\bigg{(}\omega_{\mu\nu\rho}- {1\over 2} H_{\mu\nu\rho}\sigma_3
\bigg{)}
\Gamma^{\mu\nu}+\bigg{(}{1\over 48}F_{\mu\nu\eta}\Gamma^{\mu\nu\eta}
\sigma_1\cr && \cr &&\;\;\;\;\;\;\;\;\;\;\;\;\;\;\;\;\;\;+
{1\over 480} F_{\mu\nu\lambda\delta\eta}
\Gamma^{\mu\nu\lambda\delta\eta}\sigma_0\bigg{)}\Gamma_{\rho}\bigg{]},
\label{ACF222}
\eea
for type IIB string theory \footnote{Here we have fixed the numerical 
coefficient of the RR fields as those in \cite{Warner}.}.

In the light-cone gauge we set $x^+=p^+\tau,\;\Gamma^+\theta^p=0$,
 then the non-zero contribution to the (\ref{FAC}) comes only from
the term where both the ``external'' and ``internal'' $\partial_ax^{\mu}$
factors become $p^+\delta_+^{\mu}\delta_a^{0}$. Therefore the action in
the light-cone gauge reads
\be
-{ip^+\over\pi}\int
d\sigma^2\;{\bar\theta}^p\Gamma_+\bigg{(}\delta_{pq} {\cal
D}_{\tau}+(\sigma_3)_{pq} {\cal D}_{\sigma}\bigg{)}\theta^q,
\label{ACF1}
\ee
where the supercovariant derivatives are given by
\bea
{\cal D}_{\tau}&=&\partial_{\tau}+{1\over 4}p^+ \left[(\omega_{\mu\nu
+}- {1\over 2} H_{\mu\nu +}\sigma_3) \Gamma^{\mu\nu}+({1\over 2}F_{\mu\nu}
\Gamma^{\mu\nu}\sigma_0+{1\over
48} F_{\mu\nu\lambda\delta}
\Gamma^{\mu\nu\lambda\delta}\sigma_1) \Gamma_{+}\right],\cr &&\cr
{\cal D}_{\sigma}&=&\partial_{\sigma}\;,
\label{SUCO}
\eea
for type IIA string theory. Here $\sigma_i$'s are Pauli matrices.
Similarly one can write the supercovariant derivatives for type IIB
in the light-cone gauge. We note that in our case the gravity solutions
(\ref{rrrr2}) have just one RR field which gives a mass terms for fermions.

As we see the fermion mass comes from the term corresponding to the 
RR $(p+1)$-form field. One can simply write down the equations of motion and 
solve them, though we will not study the fermionic solution in this paper. 

As a particular example one can consider the case of $p=5$. The
background (\ref{rrrr}) will then provide the supergravity description of
``New'' six-dimensional gauge theory \cite{Witten} (see also \cite{Kol}). 
The supergravity description of new gauge theory has been studied in \cite{AO}.
While we were typing our paper we received paper \cite{OS} where this
model has been studied in detail.


\section{Penrose limit of NCOS in four dimensions}

The worldvolume theory of D-brane in the presence of B field with one
leg along the time direction (E field) is an interacting theory which 
includes open string. These theories are known as noncommutative open 
string theory or in short NCOS \cite{OM}. The supergravity description 
of these theories have been studied in \cite{all}. The corresponding 
gravity solution for NCOS in four dimensions is given by \cite{GMMS}
\bea
ds^2&=&R^2h^{1/2}\bigg{[}r^2\left(-dt^2+dx_1^2+h^{-1}(dx_2^2+dx_3^2)\right)
+{dr^2\over r^2}+d\Omega^2_5\bigg{]}\cr
e^{2\phi}&=&g_s h,\;\;\;\;\;\;\;\;\;\;h=1+a^4r^4\cr
B_{01}&=&a^4r^4,\;\;\;\;\;A_{23}={a^4r^4\over 1+a^4r^4},
\label{NCOS}
\eea
where $a^4\sim {b\over {\tilde g}_s N}$ with $b$ is the noncommutative 
parameter and ${\tilde g}_s$ is effective string coupling appears in 
the theory after taking the decoupling limit. 

The NCOS$_4$ is S-dual to NCSYM$_4$ \cite{GMMS}. Therefore the corresponding 
supergravity solutions also map to each other under S-duality. The Penrose 
limit of NCSYM$_4$ has been studied in \cite{HRV}, the 
resulting PP-wave metric in
the Brinkman form of NCSYM$_4$ is given by 
\bea
ds^2&=&2dudv-l^2(x^2+z^2+y_1^2+F y_2^2)du^2\cr
&+& dx^2+d{\vec z}^2+dy_1^2+dy_2^2\;,
\label{NCPP}
\eea
where $F$ is a known function of $u$, for which we do not need 
an explicit form in our consideration.

According to the proposal of \cite{BMN}, there must 
be a subsector of NCSYM$_4$ theory
which describes type IIB string theory in this PP-wave background. Of course,
neither this background may be a solvable one for type IIB, 
nor the corresponding gauge theory description may be tractable.
Nevertheless one could ask if the relation between NCOS$_4$ and 
NCSYM$_4$ is still the same after Penrose limit. In other words, if there is 
a subsector of NCOS$_4$ describing type IIB string theory in a PP-wave
background coming from the Penrose limit of (\ref{NCOS})
which maps to the subsector of NCSYM$_4$ describing type IIB string theory
on the PP-wave background (\ref{NCPP}).

We note, however, that we can take several null geodesic each 
giving a PP-wave in both NCSYM and NCOS. 
But starting with a null geodesic in NCSYM theory, 
one can find only one null geodesic in NCOS which map together 
under S-duality. These two Penrose limits then give subsectors of 
NCSYM and NCOS theories which
map togther under S-duality. To demonstrate this procedure let us consider 
the Penrose limit of (\ref{NCOS}).

Consider a null geodesic in the $(t,r,\psi)$ plane at the fixed point with
respect to other coordinates. This geodesic is generated by tangent vector 
${\dot t}\partial_t+{\dot r}\partial_r+{\dot \psi}\partial_{\psi}$, where
dot denotes derivative  with respect to the affine parameter.
Since $\partial_t$ and $\partial_{\psi}$ are killing vectors, they define
constant of motion along geodesic, in fact
\be
E=h^{1/2}r^2{\dot t},\;\;\;\;\;\;J=h^{1/2}{\dot \psi}
\ee
are conserved quantities corresponding to the energy and angular momentum.
For a null geodesic we get
\be
{\dot r}^2={1\over h}-{r^2l^2\over h},
\ee
where $l={J\over E}$ and we have also rescaled the affine parameter by $E$.

Consider a coordinate transformation from $(t,r,\psi)\rightarrow
(u,v,x)$, which is more suitable coordinate to write the metric. The 
coordinate transformation is given by:
\bea
dr&=& \left({1\over h}-{r^2l^2\over h}\right)^{1/2}du\;,\cr &&\cr
dt&=&{1\over r^2h^{1/2}}du-dv+ldx\;,\cr &&\cr
d\psi&=&{l\over h^{1/2}}du+dx\;.
\eea
Substituting this change of coordinate in the metric 
we get
\bea
ds^2&=&R^2\bigg{[}2dudv-{l^2\over h^{1/2}}\sin^2\theta\;du^2
-2l\sin^2\theta\;dudx+h^{1/2}
(\cos^2\theta-l^2r^2)dx^2\cr &\cr
&&\;\;\;\;\;\;\;\;+\;2lr^2h^{1/2}
dvdx-r^2h^{1/2}dv^2+
r^2h^{1/2}dw^2\cr &&\cr
&&\;\;\;\;\;\;\;\;+\;r^2h^{-1/2}(dy^2+
y^2\; d\alpha^2)+h^{1/2}(d\theta^2+
\sin^2\theta^2\;d\Omega_3^2)\bigg{]}\;
\eea
Note that here we have also changed the coordinates as 
\bea
dx_1^2 & \rightarrow &
dw^2\;,\cr
dx_2^2+dx_3^2 & \rightarrow &
dy^2+y^2\; d\alpha^2\;.
\eea
We now rescale the coordinate as following
\be
u\rightarrow u\;,\;\;\;v\rightarrow{v\over R^2}\;,\;\;\;
\theta\rightarrow
{z\over R}\;,\;\;\;\;x\rightarrow{x\over R}\;,\;\;\;\;
w\rightarrow {w\over R}\;,
\;\;\;\; y\rightarrow {y\over R}\;,
\ee 
and all other coordinates remain fixed. Taking the Penrose limit in which
$R\rightarrow \infty$ and keeping $a$ fixed, we find the PP-wave  
metric of NCOS$_4$ as following
\bea
ds^2&=&2dudv-{l^2\over h^{1/2}}z^2du^2+h^{1/2}(1-l^2r^2)dx^2
+h^{1/2}
(dz^2+z^2d\Omega^3_2)\cr &&\cr &+&
r^2h^{1/2}dw^2+r^2h^{-1/2}(dy^2+y^2d\alpha^2)
\label{NCOSR}
\eea
in this metric $h$ is a function of $u$. This is the PP-wave in 
Rosen coordinate.
To find the PP-wave in  Brinkman we use the following transformation
\be
u\rightarrow u,\;\;\;\;x\rightarrow {x\over h^{1/4}\sqrt{1-l^2r^2}},
\;\;\; z\rightarrow {z\over h^{1/4}},\;\;\;w\rightarrow {w\over r h^{1/4}},
\;\;\;y\rightarrow {h^{1/4}y\over r}\;,
\ee
and 
\be
v\rightarrow v+ {1\over 2}\frac{\partial_u( h^{1/4}\sqrt{1-l^2r^2})}
{h^{1/4}\sqrt{1-l^2r^2}} x^2+
{1\over 2}\frac{\partial_u h^{1/4}}{h^{1/4}} z^2+
{1\over 2}\frac{\partial_u (r h^{1/4})}{r h^{1/4}} w^2+
{1\over 2}\frac{\partial_u (r /h^{1/4})}{r/h^{1/4}} y^2.
\ee
Using this transformation the metric reads
\be
ds^2=2dudv+(F_w w^2+F_x x^2+F_y y^2+F_z z^2)du^2+dw^2+dx^2+
d{\vec y}^2+d{\vec z}^2,
\ee
where
\be\ba {ll}
F_w=\frac{\partial_u^2 (r h^{1/4})}{r h^{1/4}}\;,&
F_y=\frac{\partial_u^2 (r/ h^{1/4})}{r/ h^{1/4}}\;,
\cr & \cr
F_z=-{l^2\over h}+\frac{\partial_u^2  h^{1/4}}
{ h^{1/4}}\;,
&
F_x=\frac{\partial_u^2  (h^{1/4}
\sqrt{1-l^2r^2})}
{ h^{1/4}\sqrt{1-l^2r^2}}\;,
\ea\ee
and one can show that $F_w = F_z = F_x$.

We note, however, that one could also consider the Penrose limit of the 
theory for case where the noncommutative parameters is kept fixed in the
limit of $N\rightarrow \infty$. In this case $a\rightarrow 0$, thus
$h\rightarrow 1$. Therefore the PP-wave solution we get is exactly
the one coming from $AdS_5\times S^5$. 

Under S-duality, the metric in string frame changes as $ds^2\rightarrow 
e^{-\phi} ds^2$, and therefore performing S-duality on 
the PP-wave solution (\ref{NCOSR}) one gets  
\bea
ds^2&=&2h^{-1/2}dudv-{l^2\over h }z^2du^2+(1-l^2r^2)dx^2
+(dz^2+z^2d\Omega^3_2)\cr &&\cr &+&
r^2dw^2+r^2h^{-1}(dy^2+y^2d\alpha^2)\;.
\eea
On the other hand under S-duality the null geodesic is also changed such that
the affine parameter of the null geodesic changes as 
$du\rightarrow h^{1/2} du$.
Using this map the above PP-wave solution reads
\bea
ds^2&=&2dudv-l^2z^2du^2+(1-l^2r^2)dx^2
+(dz^2+z^2d\Omega^3_2)\cr &&\cr &+&
r^2dw^2+r^2h^{-1}(dy^2+y^2d\alpha^2)\;,
\eea
which is the PP-wave metric of NCSYM$_4$ in the Rosen coordinates,
and moreover goes to 
the same form as (\ref{NCPP}) in the Brinkman coordinates. Note also that the
null geodesic equation also maps to ${\dot r}^2=1-l^2r^2$.

More generally for a PP-wave 
solution with string metric in the Rosen coordinates as
\be
ds^2=2dudv+G(u)dx^2,
\ee
one can write the solution in the Brinkman form as following
\be
ds^2=2dudv+{1\over 2}x^2F(G) du^2+dx^2\;,
\ee
where F(G) is a function of $G$ and its derivative. Under S-duality the metric 
changes to
\be
ds^2=2dudv+{1\over 2}x^2{\tilde F}(G) du^2+dx^2\;,
\ee
where 
\be
{\tilde F}(G)=F(G)+{1\over 2}(\partial_u \phi)^2-
\frac{\partial_u (G\partial_u\phi)}{G}\;,
\ee
here $\phi$ is the dilaton.

We therefore conclude that under S-duality the subsector of NCOS$_4$ describing
type IIB string theory in a PP-wave
background, coming from the Penrose limit of (\ref{NCOS}),
maps to a subsector of NCSYM$_4$ describing type IIB string theory
on the PP-wave background (\ref{NCPP}). Note that under S-duality the 
corresponding null geodesics are also map to each other. In fact only for these
geodesics one can compare the corresponding subsectors. The Penrose limit
of other NCOS can also be done in the same as presented in this section.

\section{Penrose limit of OM theory}

The worldvolume theory of M5-brane in the presence of C-field is decoupled from
the bulk gravity leading to a nontrivial interacting theory with light open 
membrane. This theory has been introduced in \cite{OM} and called OM-theory. 
Starting with $N$ M5-branes on top of each other with a nonzero C-field 
parametrized by $b$, at the decoupling limit we find the gravity solution 
corresponding to OM-theory as following \cite{AOS}
\bea
l_p^{-2}ds^2&=&4R^2h^{1/3}\bigg{[}r^2\left(-dt^2+dx_1^2+dx_2^2+h^{-1}
(dx_3^3+dx_4^2+dx_5^2)\right)+{dr^2\over r^2}\cr &&\cr
&+&{1\over 4}\left(d\theta^2+\cos^2\theta\; d\psi^2+\sin^2\theta\; 
d\Omega_2^2\right)\bigg{]}
\cr &&\cr
h&=&1+{b^3\over R^3}r^6,\;\;\;\;\;dC=l_p^3N \;\epsilon_4\;,\cr &&\cr
C_{012}&=&l_p^3
{ b^{3/2}\over R^3}r^6,\;\;\;\;\;\;\;C_{345}=l^3_p
{ b^{3/2}\over R^3}\frac{r^6}{1+{b^3\over R^3}r^6}\;,
\label{OMSOL}
\eea
where $\epsilon_4$ is the volume of $S^4$ part of the above metric.

In this section we would like to consider the Penrose limit of OM theory.
To do this, in the same way as in the previous sections, 
let us consider a null geodesic 
in the $(t,r,\psi)$ plane at the fixed point with
respect to other coordinates. This geodesic is generated by tangent vector 
${\dot t}\partial_t+{\dot r}\partial_r+{\dot \psi}\partial_{\psi}$, where
dot denotes derivative  with respect to the affine parameter.
Since $\partial_t$ and $\partial_{\psi}$ are killing vectors, they define
constant of motion along geodesic, in fact
\be
E=h^{1/3}r^2{\dot t},\;\;\;\;\;\;J={h^{1/3}\over 4}{\dot \psi}
\ee
are conserved quantities corresponding to the energy and angular momentum.
For a null geodesic we get
\be
{\dot r}^2={1\over h^{2/3}}-{4r^2l^2\over h^{2/3}},
\ee
Now consider a coordinate transformation from $(t,r,\psi)\rightarrow
(u,v,x)$ which is more situable coordinate to write the metric. The 
coordinate transformation is given by
\bea
dr&=& \left({1\over h^{2/3}}-{4r^2l^2\over h^{2/3}}\right)^{1/2}du\;,\cr &&\cr
dt&=&{1\over r^2h^{1/3}}du-dv+ldx\;,\cr &&\cr
d\psi&=&{4l\over h^{1/3}}du+dx\;.
\label{COOR}
\eea
One can use the first equation to find $r$ as a function of $u$
\be
u=\int {h^{1/3}dr\over \sqrt{1-4r^2l^2}}
\ee 
which could be inverted to find $r(u)$. Then we can plug this to the
other expression in the metric to write the metric (\ref{OMSOL}) in 
terms of the $u$. Substituting this change of coordinate in the metric 
(\ref{OMSOL}),
and rescaling the coordinate as following:
\be
u\rightarrow u\;,\;\;\;v\rightarrow{v\over R^2}\;,\;\;\;\theta\rightarrow
{z\over R}\;,\;\;\;\;x\rightarrow{x\over R}\;,\;\;\;\;
w\rightarrow {w\over R}\;,
\;\;\;\; y\rightarrow {y\over R}\;,
\ee 
with all other coordinates remaining fixed, we obtain by taking the 
Penrose limit ($R\rightarrow \infty$ while ${b\over R}$ fixed), 
the PP-wave  metric of OM theory in Rosen coordinate as following:
\bea
ds^2&=&2dudv-{4l^2\over h^{1/3}}z^2du^2+
h^{1/3}({1\over 4}-l^2r^2)dx^2+{h^{1/3}\over 4}
(dz^2+z^2d\Omega^2_2)\cr &&\cr &+&
r^2h^{1/3}(dw^2+w^2d\alpha^2)+r^2h^{-2/3}(dy^2+y^2d{\tilde \Omega}_2^2)\;.
\eea
Note that in the above we have changed the coordinates as 
\bea
dx_1^2+dx_2^2 & \rightarrow &
dw^2+w^2\;d\alpha^2\;,\cr
dx_3^2+dx_4^3+dx_5^2 & \rightarrow &
dy^2+y^2\; d{\tilde \Omega}_2^2\;.
\eea
In the same way, as in previous sections, one can write the PP-wave in the
Brinkman coordinates which is: 
\be
ds^2=2dudv+(F_w w^2+F_x x^2+F_y y^2+F_z z^2)du^2+d{\vec w}^2+dx^2+
d{\vec y}^2+d{\vec z}^2,
\label{om-g1}
\ee
where
\be\ba {ll}
F_w=\frac{\partial_u^2 (r h^{1/6})}{r h^{1/6}}\;,&
F_y=\frac{\partial_u^2 (r/ h^{1/3})}{r/ h^{1/3}}\;,
\cr & \cr
F_z=-{16l^2\over h^{2/3}}+\frac{\partial_u^2  h^{1/6}}{ h^{1/6}}\;,
&
F_x=\frac{\partial_u^2  (h^{1/6}\sqrt{{1\over 4}-l^2r^2})}
{ h^{1/6}\sqrt{{1\over 4}-l^2r^2}}\;.
\label{om-g2}
\ea\ee
One can show once again that $F_w = F_z = F_x$.
We also note that in this limit we get only one C field corresponding
to the $N$ M5-branes as following
\be
(dC)_{u,\vec{z}}=32 l^3_p h^{-5/6},
\ee
all others go to zero in the Penrose limit. 

Note that in getting the above PP-wave we have assumed that the noncommutative
parameter $b$ is also going to infinity in large $N$ limit. But we could
consider the case in which the noncommutative parameter remains fixed in the
Penrose limit. In this case $h\rightarrow 1$ and then we just get maximally
supersymmetric 11-dimensional PP-wave. 

\section{Conclusions}

In this paper we have studied the PP-wave limit of ODp, NOCS and 
OM theories. We
have shown that there are two different Penrose limits one can take for
ODp theories in which one of them leads to an exactly solvable theory. The 
metric of this PP-wave is very similar to that considered by Nappi-Witten    
\cite{NW}. But, besides the NS B field we also have an RR field in 
this background. This R-NS PP-wave also gives an exact solution 
of worldsheet string 
theory. We have studied the light-cone GS action for this model.

We have shown that under S-duality the subsector of 
NCOS$_4$ which is dual to type IIB string theory on PP-wave of NCOS, maps to
the subsector of NCSYM$_4$, dual to type IIB string theory on PP-wave
of NCSYM. In fact, this statement is true only for those null
geodesics which are S-dual to each other.

An interesting fact about the Penrose limit is that it can wash away the
noncommutativity effects. More precisely, if we assume that the 
noncommutativity parameter is kept fixed in taking the Penrose limit, the
PP-wave solution we get is the same as if we had considered the Penrose limit
of the corresponding commutative theory. According to BMN \cite{BMN} proposal,
this means, for example, that type IIB string theory on 
maximally supersymmetric PP-wave
background has a gauge theory description in terms of a subsector of
non-commutative SYM theory in four dimensions with 16 supercharges. We however
leave the detail of this correspondence for further study.    

\vspace{.5cm}

{\bf Acknowledgments}

We would like to thank Yaron Oz for e-mail correspondence and for bringing 
our attentions to the correct expressions for RR-fields in ODp-theory.


\begin{thebibliography}{99}

\bibitem{metsaev}
R.R.Metsaev, ``Type IIB Green-Schwarz superstring in plane wave 
Ramond-Ramond background,'' {\em Nucl.Phys.} {\bf B625} (2002) 70, 
hep-th/0112044.

 R.R. Metsaev and A.A. Tseytlin, 
``Exactly solvable model of superstring in Ramond-Ramond 
plane wave background,'' {\em Phys.Rev.} {\bf D65} (2002) 
126004, hep-th/0202109.


\bibitem{hull}
M.~Blau, J.~Figueroa-O'Farrill, C.~Hull and G.~Papadopoulos,
``Penrose limits and maximal supersymmetry,''
{\em Class. Quant. Grav.}  {\bf 19}, L87 (2002), 
hep-th/0201081.


\bibitem{blau}  
M.~Blau, J.~Figueroa-O'Farrill and G.~Papadopoulos,
``Penrose limits, supergravity and brane dynamics,''
,hep-th/0202111.


\bibitem{BMN}
D. Berenstein, J. Maldacena and H. Nastase, ``String
in flat space and pp-waves from ${\cal N}=4$ Super
Yang Mills,'' hep-th/0202021.

\bibitem{all2}
M. Blau, J. Figueroa-O'Farrill and G. Papadopoulos,``Penrose
limits, supergravity and brane dynamics,'' hep-th/0202111.

M. Cvetic, H. Lu and C.N. Pope, ``Penrose Limits, PP-Waves
and Deformed M2-branes,''
hep-th/0203082.

R. Gueven, ``Randall-Sundrum Zero Mode as a Penrose Limit,''
{\em Phys.Lett.} {\bf B535} (2002) 309, hep-th/0203153.

S. R. Das, C. Gomez and S-J. Rey, ``Penrose limit,
Spontaneous Symmetry Breaking and Holography in PP-Wave Background,''
hep-th/0203164.

M. Cvetic, H. Lu and C.N. Pope, ``M-theory PP-waves, Penrose
Limits and Supernumerary Supersymmetries,'' hep-th/0203229.


H. Lu and J.F. Vazquez-Poritz, ``Penrose Limits of Non-standard
Brane Intersections,''
hep-th/0204001.


A. Parnachev and D. A. Sahakyan, ``Penrose limit and string
quantization in $AdS_3 \times S^3$,'' hep-th/0205015.


S. Seki, ``D5-brane in Anti-de Sitter Space and Penrose Limit,''
hep-th/0205266.

D. Mateos and S. Ng, ``Penrose Limits of the Baryonic D5-brane,''
hep-th/0205291.

E. G. Gimon, L. A. Pando Zayas and J. Sonnenschein, ``Penrose
Limits and RG Flows,'' hep-th/0206033.

A. Feinstein, ``Penrose Limits, the Colliding Plane Wave Problem
and the Classical String Backgrounds,'' hep-th/0206052.

D. Mateos, ``Penrose Limits, Worldvolume Fluxes and Supersymmetry.''
hep-th/0206194.

A.A. Zheltukhin and D.V. Uvarov, ``An Inverse Penrose Limit
and Supersymmetry Enhancement in the Presence of Tensor
Central Charges,'' hep-th/0206214.


\bibitem{all1}
N. Itzhaki, I. R. Klebanov and S. Mukhi, ``
PP Wave Limit and Enhanced Supersymmetry in Gauge
Theories,'' {\em JHEP} {\bf 0203} (2002) 048,
hep-th/0202153.

J. Gomis and H. Ooguri, ``Penrose Limit of $N=1$ Gauge
Theories,'' hep-th/0202157.

L. A. Pando Zayas and J. Sonnenschein, ``On Penrose Limits
and Gauge Theories,'' {\em JHEP} {\bf 0205} (2002) 010,
hep-th/0202186.

M. Alishahiha and M. M. Sheikh-Jabbari, ``The PP-Wave
Limits of Orbifolded $AdS_5\times S^5$,'' {\em Phys.Lett.}
{\bf B535} (2002) 328, hep-th/0203018.

N. Kim, A. Pankiewicz, S-J. Rey and S. Theisen,
``Superstring on PP-Wave Orbifold from Large-N
Quiver Gauge Theory,'' hep-th/0203080.

T. Takayanagi and S. Terashima, ``Strings on
Orbifolded PP-waves,'' hep-th/0203093.

U. Gursoy, C. Nunez and M. Schvellinger, ``RG flows from
Spin(7), CY 4-fold and HK manifolds to AdS, Penrose limits
and pp waves,'' hep-th/0203124.


E. Floratos and A. Kehagias, ``Penrose Limits of
Orbifolds and Orientifolds,'' hep-th/0203134.


S. Mukhi, M. Rangamani and E. Verlinde, ``Strings
from Quivers, Membranes from Moose,'' {\em JHEP}
{\bf 0205} (2002) 023, hep-th/0204147.

M. Alishahiha and M. M. Sheikh-Jabbari, ``Strings in
PP-Waves and  Worldsheet Deconstruction,'' {\em Phys. Lett.}
 {\bf B538} (2002) 180, hep-th/0204174.

K. Oh and R. Tatar, ``Orbifolds, Penrose Limits and
Supersymmetry Enhancement,'' hep-th/0205067.

Y. Hikida and Y. Sugawara. ``Superstrings on PP-Wave
Backgrounds and Symmetric Orbifolds, hep-th/0205200.

C. Ahn, ``More on Penrose Limit of $AdS_4 \times
Q^{1,1,1}$,'' hep-th/0205008.

C. Ahn, ''Comments on Penrose Limit of $AdS_4
\times M^{1,1,1}$,'' hep-th/0205109.

C. Ahn, ``Penrose Limit of $AdS_4\times V_{5,2}$ and
Operators with Large R Charge,''
hep-th/0206029.

S. G. Naculich, H. J. Schnitzer and N. Wyllard, ``pp-wave limits
and orientifolds,'' hep-th/0206094.

\bibitem{HRV}
V. E. Hubeny, M. Rangamani and E. Verlinde, ``Penrose Limits and 
Non-local theories,'
hep-th/0205258.


\bibitem{Berkooz}
M. Berkooz, M. Rozali and N. Seiberg, ``Matrix Description of 
M-theory on $T^4$ and $T^5$,'' {\em Phys.Lett.}
{\bf B408} (1997) 105, hep-th/9704089. 


\bibitem{Seiberg}
N. Seiberg, ``Matrix Description of M-theory on $T^5$ 
and $T^5/Z_2$,'' {\em Phys.Lett.} {\bf B408} (1997) 98,
hep-th/9705221. 


\bibitem{Aharony}
O. Aharony, M. Berkooz, D. Kutasov and N. Seiberg, ``
Linear Dilatons, NS5-branes and Holography,'' 
{\em JHEP} {\bf 9810} (1998) 004, hep-th/9808149. 


\bibitem{Hashimato}
A. Hashimoto and N. Itzhaki, ``Non-Commutative Yang-Mills and 
the AdS/CFT Correspondence,'' {\em Phys.Lett.} {\bf B465} (1999) 
142, hep-th/9907166. 


\bibitem{MalRU}
J. M. Maldacena and J. G. Russo, ``Large N Limit of Non-Commutative 
Gauge Theories,'' {\em JHEP} {\bf 9909} (1999) 025, hep-th/9908134. 


\bibitem{OM}
R. Gopakumar, S. Minwalla, N. Seiberg and A. Strominger, ``
OM Theory in Diverse Dimensions,'' {\em JHEP} {\bf 0008} 
(2000) 008, hep-th/0006062.


\bibitem{GMMS}
R. Gopakumar, J. Maldacena, S. Minwalla and A. Strominger, ``
S-Duality and Noncommutative Gauge Theory,'' {\em JHEP}
{\bf 0006} (2000) 036, hep-th/0005048.
 
\bibitem{HARM1}
T. Harmark, ``Open Branes in Space-Time Non-Commutative 
Little String Theory,'' {\em Nucl.Phys.} {\bf B593} 
(2001) 76, hep-th/0007147. 

\bibitem{AOS}
M, Alishahiha, Y. Oz and M.M. Sheikh-Jabbari,``Supergravity and Large 
N Noncommutative Field Theories,''{\em JHEP} {\bf 9911} (1999) 007;
hep-th/9909215.


\bibitem{all}
T. Harmark, ``Supergravity and Space-Time Non-Commutative 
Open String Theory,'' {\em JHEP} {\bf 0007} (2000) 043;
hep-th/0006023.


\bibitem{ALI1}
M. Alishahiha, ``On Type II NS5-branes in the presence of an RR 
field,'' {\em Phys.Lett.} {\bf B486}(2000) 194, hep-th/0002198

 
\bibitem{AOR}
M. Alishahiha, Y. Oz and J. G. Russo, ``Supergravity and 
Light-Like Non-commutativity,'' {\em JHEP} {\bf 0009} (2000) 002;
hep-th/0007215. 


\bibitem{mitra-roy}
I. Mitra and S. Roy,``(NS5,Dp) and (NS5,D(p+2),Dp) bound states 
of type IIB and type IIA string theories ''  
{\em JHEP} {\bf 0102} (2001) 026; hep-th/0011236.



\bibitem{all5}
M. Li and Y-S. Wu, ``Holography and Noncommutative Yang-Mills,''
{\em Phys.Rev.Lett.} {\bf 84} (2000) 2084;
hep-th/9909085. 

T. Harmark and N.A. Obers, ``Phase Structure of Non-Commutative 
Field Theories and Spinning Brane Bound States,'' {\em JHEP}
{\bf 0003} (2000) 024; hep-th/9911169.

J.L.F. Barbon and E. Rabinovici, ``On 1/N Corrections to the
Entropy of Noncommutative Yang-Mills Theories,'' {\em JHEP}
{\bf 9912} (1999) 017; hep-th/9910019.

J. X. Lu and S. Roy, ``(p + 1)-Dimensional Noncommutative Yang-Mills
and D(p - 2) Branes,'' {\em Nucl.Phys.} {\bf B579} (2000) 229; 
hep-th/9912165. 

R.-G. Cai and N. Ohta, ``Noncommutative and Ordinary Super 
Yang-Mills on (D$(p-2)$, D$p$) Bound States,'' {\em JHEP}
{\bf 0003} (2000) 009;
hep-th/0001213. 

D. S. Berman, V. L. Campos, M. Cederwall, U. Gran, H. Larsson, 
M. Nielsen, B. E.W. Nilsson and P. Sundell, ``Holographic 
Noncommutativity,'' {\em JHEP} {\bf 0105} (2001) 002, 
hep-th/0011282.

\bibitem{RG}
E. G. Gimon, L. A. Pando Zayas, J. Sonnenschein, 
``Penrose Limits and RG Flows'', hep-th/0206033.

\bibitem{RG1}
Richard Corrado, Nick Halmagyi, Kristian D. Kennaway, Nicholas P. Warner,
``Penrose Limits of RG Fixed Points and PP-Waves with Background
Fluxes'', hep-th/0205314.


\bibitem{CAS}
R. Casero, ``Penrose limit of a non-supersymmetric RG fixed point'', 
hep-th/0207221.


\bibitem{NW}
C.R. Nappi and E. Witten, ``A WZW model based on a 
non-semi-simple group,'' {\em Phys.Rev.Lett.} {\bf 71}
(1993) 3751, hep-th/9310112.


\bibitem{Amati}
A. Amati and C. Klimcik, ``Nonperturbative Computation Of 
The Weyl Anomaly For A Clas of Nontrivial Backgrounds.''
{\em Phys.Lett.} {\bf B219} (1989) 443. 


\bibitem{HOR}
G.T. Horowitz and A.R. Steif, ``Space-Time Singularities 
In String Theory,'' {\em Phys.Rev.Lett.} {\bf 64} (1990) 260. 


\bibitem{KP}
E. Kiritsis and B. Pioline, ``Strings in homogeneous gravitational 
waves and null holography,'' hep-th/0204004.
 


\bibitem{RT}
J.G. Russo and A.A. Tseytlin, ``On solvable models of type IIB
superstring in NS-NS and R-R plane wave backgrounds,''
{\em JHEP} {\bf 0204} (2002) 021, hep-th/0202179.


\bibitem{Warner}
R. Corrado, N. Halmagyi, K. D. Kennaway and N. P. Warner, ``Penrose
Limits of RG Fixed Points and PP-Waves with Background Fluxes,''
hep-th/0205314.

\bibitem{Myers}
D. Brecher, C. V. Johnson, K. J. Lovis and R. C. Myers, ``Penrose
Limits, Deformed pp-Waves and the String Duals of N=1 Large n
Gauge Theory,'' hep-th/0206045.

\bibitem{all6}
A. Dabholkar and S. Parvizi, ``Dp Branes in PP-wave Background,''
hep-th/0203231.

A. Kumar, R. R. Nayak and Sanjay, ``D-Brane Solutions in pp-wave 
Background,'' hep-th/0204025.


A. Parnachev and D. A. Sahakyan, ``Penrose limit and string 
quantization in $AdS_3 \times S^3$,'' {\em JHEP} {\bf 0206}
(2002) 035, hep-th/0205015. 


Y. Hikida and Y. Sugawara, ``Superstrings on PP-Wave 
Backgrounds and Symmetric Orbifolds,'' 
{\em JHEP} {\bf 0206} (2002) 037, hep-th/0205200.
 


Y. Michishita, ``D-branes in NSNS and RR pp-wave backgrounds 
and S-duality,'' hep-th/0206131. 

\bibitem{OS} 
Y. Oz and T. Sakai, 
``Penrose Limit and Six-Dimensional Gauge Theories,'' hep-th/0207223.


\bibitem{AGGP}
M. Alishahiha, M. A. Ganjali, A. Ghodsi and S. Parvizi,
``On Type IIA String Theory on the PP-wave Background,''
hep-th/0207037.

\bibitem{ROM}
L. J. Romans, ``Massive N=2a Supergravity In Ten-Dimensions,''
{\em Phys.Lett.} {\bf B169} (1986) 374.

\bibitem{Hassan}
S. F. Hassan, ``T-Duality, Space-time Spinors and R-R Fields
in Curved Backgrounds,'' {\em Nucl.Phys.} {\bf B568} (2000) 145,
hep-th/9907152.

\bibitem{Witten}
E. Witten, ``New ``Gauge'' Theories In Six Dimensions,''
{\em JHEP} {\bf 9801} (1998) 001; {\em Adv.Theor.Math.Phys.}
{\bf 2} (1998) 61, hep-th/9710065.

\bibitem{Kol}
B. Kol, ``On 6d ``Gauge'' Theories with Irrational Theta Angle,''
{\em JHEP} {\bf 9911} (1999) 017, hep-th/9711017.


\bibitem{AO}
M. Alishahiha and Y. Oz, ``Supergravity and "New" Six-Dimensional 
Gauge Theories,'' {\em Phys.Lett.} {\bf B495} (2000) 418;
hep-th/0008172.





\end{thebibliography}
\end{document}